\documentclass[conference]{IEEEtran}
\IEEEoverridecommandlockouts
\usepackage{lipsum}
\usepackage{cite}
\usepackage{amsmath,amssymb,amsfonts}
\usepackage{algorithmic}
\usepackage{graphicx}
\usepackage{textcomp}
\usepackage{xcolor}
\usepackage{hyperref}

\makeatletter % changes the catcode of @ to 11
\newcommand{\linebreakand}{%
  \end{@IEEEauthorhalign}
  \hfill\mbox{}\par
  \mbox{}\hfill\begin{@IEEEauthorhalign}
}
\makeatother % changes the catcode of @ back to 12

\def\BibTeX{{\rm B\kern-.05em{\sc i\kern-.025em b}\kern-.08em
    T\kern-.1667em\lower.7ex\hbox{E}\kern-.125emX}}

\title{DG-PPU: Dynamical Graphs based Post-processing of Point Clouds extracted from Knee Ultrasounds}

\author{
    \IEEEauthorblockN{Injune Hwang}
    \IEEEauthorblockA{\textit{University of Oxford}}
    \and
    \IEEEauthorblockN{Karthik Saravanan}
    \IEEEauthorblockA{\textit{University of Oxford}}
    \and 
    \IEEEauthorblockN{Caterina Vanelli Coralli}
    \IEEEauthorblockA{\textit{University of Oxford}}
    \and
    \linebreakand % <----- NOTE HERE, breaking after the third one!
    \IEEEauthorblockN{S Jack Tu}
    \IEEEauthorblockA{\textit{NDORMS}\\ \textit{University of Oxford} \\
                     jack.tu@ndorms.ox.ac.uk}
    \and
    \IEEEauthorblockN{Stephen J Mellon}
    \IEEEauthorblockA{\textit{NDORMS}\\ \textit{University of Oxford} \\
                     stephen.mellon@ndorms.ox.ac.uk}
}

\begin{document}

\newpage
\thispagestyle{empty} % Remove headers/footers
\begin{center}
    \vspace*{3in}% Adjust this value to vertically center the text as desired

    This paper was submitted to the IEEE International Symposium on Biomedical Imaging (ISBI).\\
    This is a preprint version and may be subject to copyright.\\[1em]
    © 2025 IEEE. Personal use of this material is permitted. However, permission to reprint/republish this material for advertising or promotional purposes or for creating new collective works for resale or redistribution to servers or lists, or to reuse any copyrighted component of this work in other works, must be obtained from the IEEE.
\end{center}
\newpage

\maketitle

%
% For example:
% ------------
%\address{School\\
%	Department\\
%	Address}
%
% Two addresses (uncomment and modify for two-address case).
% ----------------------------------------------------------
%\twoauthors
%  {A. Author-one, B. Author-two\sthanks{Some author footnote.}}
%	{School A-B\\
%	Department A-B\\
%	Address A-B}
%  {C. Author-three, D. Author-four\sthanks{The fourth author performed the work
%	while at ...}}
%	{School C-D\\
%	Department C-D\\
%	Address C-D}
%
% More than two addresses
% -----------------------
% \name{Author Name$^{\star \dagger}$  Author Name$^{\star}$  Author Name$^{\dagger}$}
%
% \address{$^{\star}$ Affiliation Number One \\
%     $^{\dagger}$}Affiliation Number Two
%

%
\begin{abstract}
Patients undergoing total knee arthroplasty (TKA) often experience \emph{non-specific anterior knee pain}, arising from abnormal patellofemoral joint (PFJ) instability. Tracking PFJ motion is challenging since static imaging modalities like CT and MRI are limited by field of view and metal artefact interference. Ultrasounds offer an alternative modality for dynamic musculoskeletal imaging. We aim to achieve accurate visualisation of patellar tracking and PFJ motion, using 3D registration of point clouds extracted from ultrasound scans across different angles of joint flexion. Ultrasound images containing soft tissue are often mislabeled as bone during segmentation, resulting in noisy 3D point clouds that hinder accurate registration of the bony joint anatomy. Machine learning the intrinsic geometry of the knee bone may help us eliminate these false positives. As the intrinsic geometry of the knee does not change during PFJ motion, one may expect this to be robust across multiple angles of joint flexion. Our dynamical graphs based post-processing algorithm (DG-PPU) is able to achieve this, creating smoother point clouds that accurately represent bony knee anatomy across different angles. After inverting these point clouds back to their original ultrasound images, we evaluated that DG-PPU outperformed manual  data cleaning done by our lab technician, deleting false positives and noise with 98.2\(\%\) precision across three different angles of joint flexion. DG-PPU is the first algorithm to automate the post-processing of 3D point clouds extracted from ultrasound scans. With DG-PPU, we contribute towards the development of a novel patellar mal-tracking assessment system with ultrasound, which currently does not exist.       
\end{abstract}


\begin{figure}[t]
\centering
   \includegraphics[width=10cm]{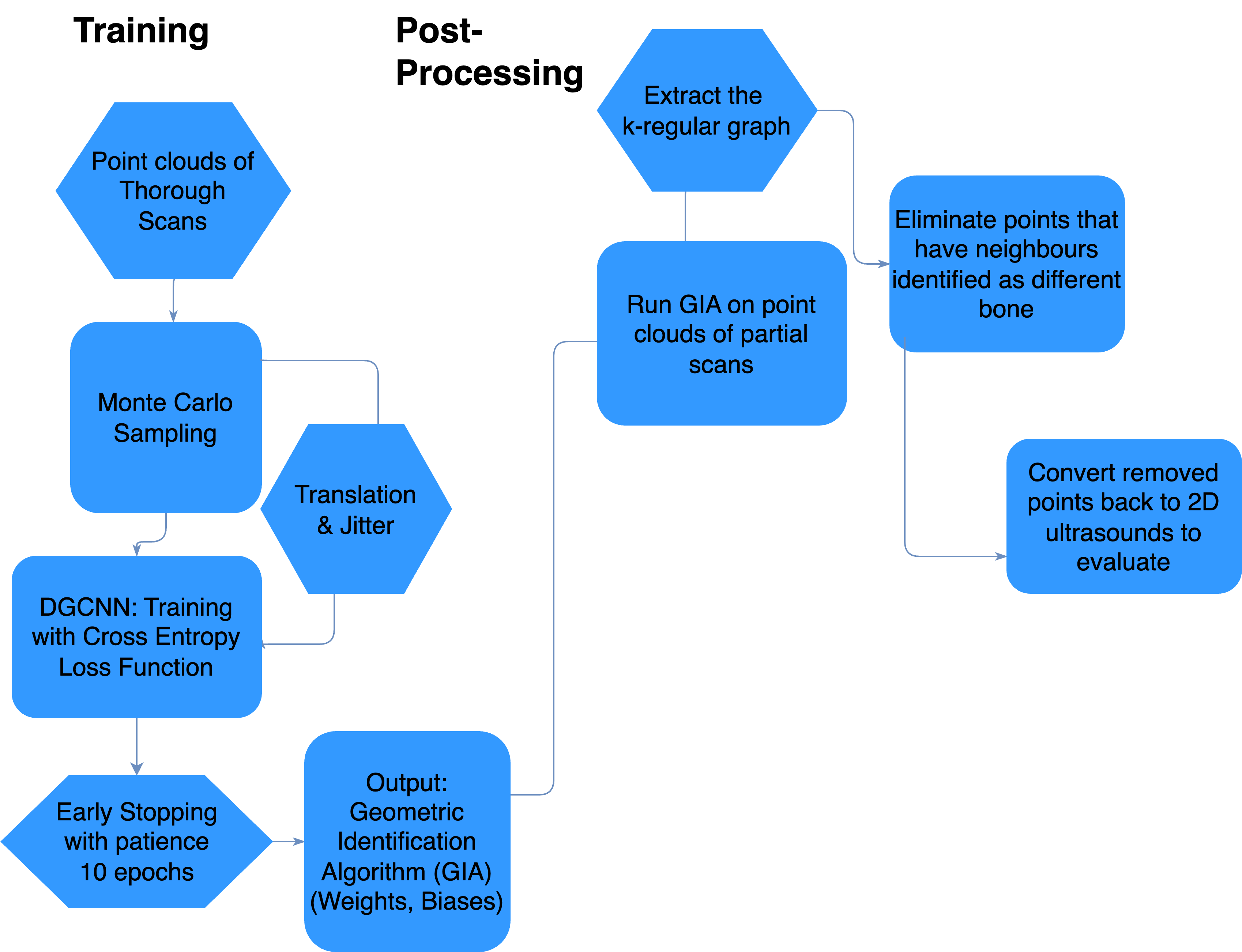}
\caption{DG-PPU}
\label{Diagram 1}
\end{figure}

\section{Introduction}

 Up to 30\% of patients report \emph{non-specific anterior knee pain} after TKA \cite{Laubach2020}. Currently, the mechanisms underlying this joint pain are not fully understood; numerous studies have previously suggested that abnormal patellofemoral instability with the trochlea of the implant is one of the main causes of this pain \cite{Petersen2013, Belvedere2014}.  Previous studies aimed to visualise this using static imaging methods such as CT and MRI \cite{Link1998}. However, consideration of the limited field view of static CT and MRI imaging, radiation exposure of CT scans, and cost of MRI imaging led to the increasing usage of ultrasound as an alternative modality for dynamic MSK imaging \cite{Nazarian2008,Khoury2007,Neustadter2004,DeSmet2002}.

Our co-authors previously developed a Computer-Aided Tracking and Motion Analysis with Ultrasound System (CATMAUS) to visualise these patellofemoral interactions \cite{MAUS,CATMAUS}.  Their most recent work was successfully able to achieve 3D point cloud reconstruction with motion tracking from 2D ultrasound scans, using a combination of semantic segmentation and point cloud registration \cite{Recent-CATMAUS}. However, the focus of this paper was on their novel tracking method---the training data they used in this study was generated from realistic plastic bone models. The long term goal is to replicate this pipeline using point clouds generated from ultrasound scans on real human bone.\\

Due to the nature of ultrasound data, point clouds extracted from 2D ultrasound images tend to be noisy with several false positives where soft tissue is mislabeled as bone. Previously, manual cleaning/de-noising was required in order to remove these erroneous points before reconstructing a 3D mesh of the joint that could be used for motion analysis of patellar tracking. In a clinical setting, this is far from ideal, as manual data cleaning cannot be done in real time to accurately visualise PFJ motion. Furthermore, the noise of the point clouds hindered the Iterative Closest Points (ICP) algorithm \cite{ICP} from finding the rigid transformations used in point cloud registration \cite{Recent-CATMAUS}. Accurate point cloud registration is what allows us to visualise 3D PFJ motion.\\

Detailed ultrasound scans are a necessity when reconstructing a realistic point cloud representation of the PFJ. To collect such data, it requires very thorough, time-consuming scanning of the joint that would not be feasible within the time frame of a standard clinic appointment. We use the term \emph{thorough scans} to describe the ultrasound scans we achieve in the lab and \emph{partial scans} for a less detailed scan one may expect to be able to use in a clinical setting. We aim to achieve accurate visualisation of PFJ motion via registration of point clouds extracted from partial scans across different angles of the knee joint. \\

With the long term goal of obtaining point cloud registration, this project focuses on de-noising and filtering the 3D point clouds extracted from the ultrasound images by CATMAUS. For this purpose, we've designed our novel algorithm: Dynamical Graph-based Post-processing for Ultrasounds (DG-PPU).  We elaborate more on the methodology we've taken below.

\section{DG-PPU}

\subsection{Dynamical Graphs Approach}
 \hspace{0.2cm} Dynamical Graph Convolutional Neural Networks (DGCNN) \cite{DynamicalGraphs} allow one to achieve classification, segmentation, and surface reconstruction tasks on point cloud data. These are tasks that a standard CNN would struggle with due to the unordered and irregular structure of point clouds \cite{ReviewiEEE}. The key idea of this approach is that the graph is updated per layer as we learn more features from the input data. Precisely, DGCNN uses a graph that connects a point to its k-nearest neighbour. As each layer learns more about the geometry, the graph would update the k-nearest neighbours accordingly. By the final epoch of training, the graph would be an accurate representation of the geometry of the point cloud as demonstrated in the test results achieved by the DGCNN when tested on ModelNet40 \cite{ModelNet40}.\\

The primary idea of DG-PPU is to use the dynamical graphs structure that is produced by the DGCNN to post-process our point clouds. The dynamical graphs at the final stage of training would let us understand the distance between two points along the knee joint.\\

Mislabelling soft tissue as bone, and including them in our point clouds, may occur due to noise accumulated from the ultrasound data itself, or due to close proximity of the ultrasound probe to neighbouring soft tissue. For the former case,  the false positive would stand out amongst a cluster of identically labelled points. For the latter, in terms of Euclidean distance, it may not be obvious. However, learning the intrinsic geometry of the knee bones from 3D point clouds allows us to distinguish the tip of two adjacent bones. False positives between two bones cause the estimation of both bones to touch at boundaries, their k-nearest neighbours would therefore not be classified as the same type of bone. This motivated the point elimination criterion of our post-processing algorithm DG-PPU which we describe in Fig.1. We describe the algorithm in more detail below.

\begin{figure*}[t]
   \centering
   \includegraphics[width=0.7\textwidth]{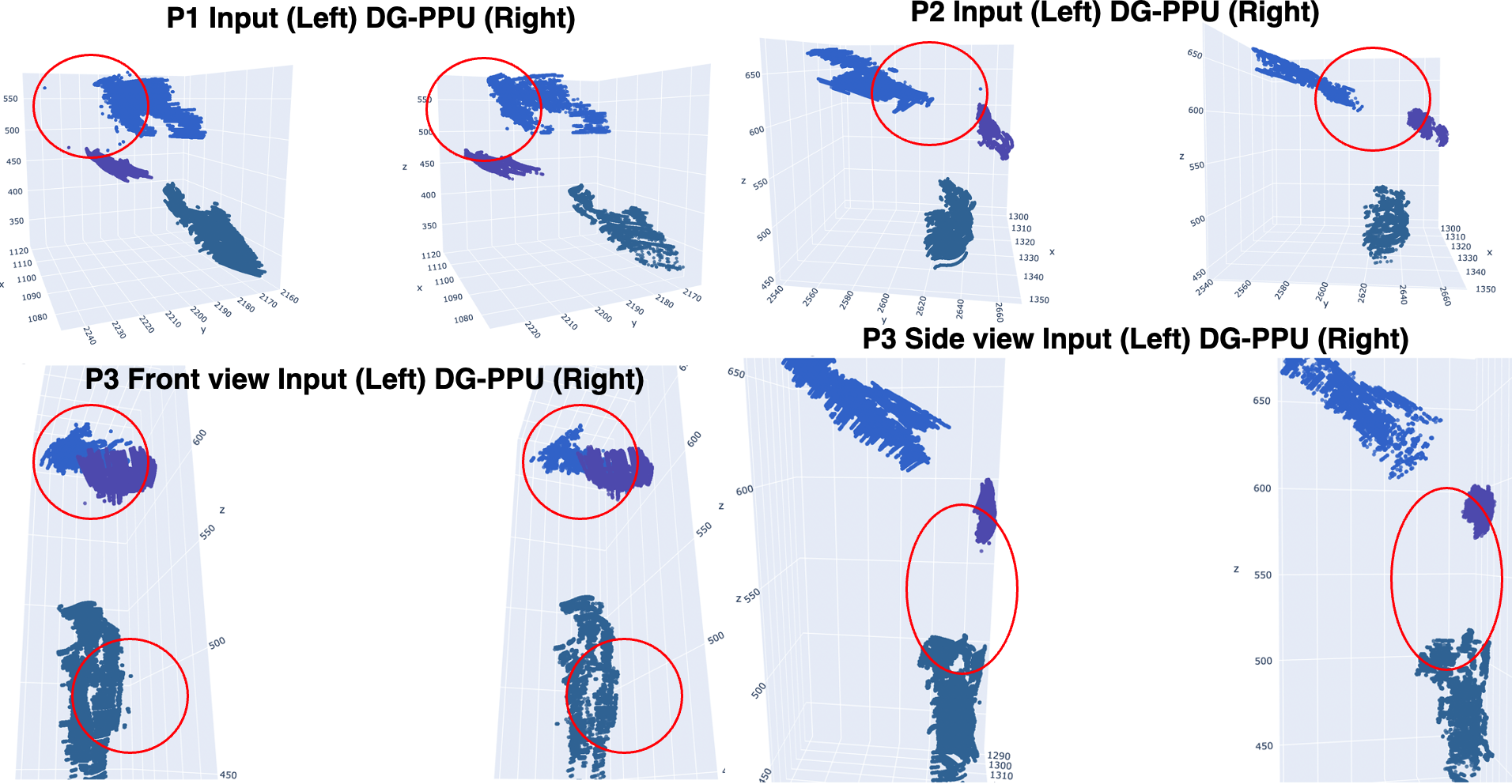} % 13cm
\caption{Post-processing results of DG-PPU on point clouds extracted from partial scans P1, P2, P3 }
\label{Diagram 3}
\end{figure*}

\subsection{Monte Carlo Sampling and Batches}

 Ultrasound scans and the point clouds we extract from them present two challenges. First, the number of points we extract from a thorough scan tend to be very large. Applying the k-regular graph \cite{RegularGraphs2023} approach to such would be computationally expensive. Second, creating a 3D point cloud reconstruction from a thorough ultrasound scan is time consuming. This motivates the usage of probabilistic models to generate batches of point clouds. We designed the algorithm to generate 500 point clouds each containing 1024 points per ultrasound scan. Batches of 1024 points make the algorithm very computationally efficient when drawing k-regular graphs. Our algorithm samples points with replacement, as we would like to represent the density accurately. As we sample 500 times (hence, 512,000 points with duplicates), even with replacement, we cover most of our point clouds when sampling.\\

Since our data set has a large imbalance in data points per bone, we implemented translation and jittering \cite{Li2020, Qi2017} to augment the minority class, doubling the number of patellar points in our dataset and minimising the likelihood that they are mislabelled by our DGCNN. We set \(k = 20\) for the k-regular graphs as a result of the following calculation. When \(k = 20\), it is of \(0.9695\) probability that we have all 500 samples have at least 20 points that represent the Patellar bone. As Patellar points are the minority class, we used them as the criterion. We could set \(k > 20\) which would allow the graphs to capture the geometry in more detail; yet, there would be a lower probability of all the samples being reasonable to use.\\  

\subsection{Training and Post-Processing}

 After we generated 500 point clouds of 1024 points per scan, we implemented DGCNN using the Pytorch Geometric python package \cite{pytorchgeo} with Adam optimizer (learning rate 0.001)\cite{Adam}, and early stopping callbacks monitoring validation loss values with 10 epochs of patience during training \cite{PausePerformance2024, KeepingInCheck2024}. Model performance metrics, including accuracy, validation loss, precision, recall, f1-score, IoU, and interactive 3D segmentation objects, were logged for each epoch during training via the wandb python package \cite{wandb}.\\

As discussed above, we trained using \(20-\)regular graphs. Each batch creates a feature point that contains geometric information updated using the graph structure per layer and epoch. The algorithm would gather the information from all these feature points to fix the weights and biases of a geometric identification algorithm. The geometric identification algorithm captures the geometric information of a point cloud we utilise to draw a \(20-\)regular graph for a general point cloud. To avoid over-fitting, we trained using scans across four different angles of knee flexion, implementing early stopping callbacks as described earlier.\\

DG-PPU operates according to the geometric identification algorithm that we've obtained above. Our algorithm searches the neighbours of each point in a batch and deletes points that have at least one neighbouring point identified as a different bone to itself. This would eliminate the false positives where soft tissue was mislabelled as bone, improving the accuracy with which we are able to visualise the PFJ and its role in patellar-tracking after position registration of these point clouds.

\section{Experiments and Results}

\subsection{Data sets and Experimental Setup}

 Our training data-set consists of 2,000 point clouds generated from thorough ultrasound scans of 4 different positions (i.e., 4 different angles of the knee joint). We denote the 4 different positions as P0, P1, P2, P3 for future reference where P0 is full flexion and P3 full extension. At full flexion, the patellar bone is distinctive; yet, for other positions, the outline of the patella becomes less evident. We test our post-processing algorithm on 1,500 point clouds generated from partial scans ranging from P1 to P3. Partial scans at a position, say P1, are sparse scans collected at the same position as its corresponding thorough scan. We test on partial scans to demonstrate that the post-processing algorithm is robust to the lack of detail and has potential clinical usage.\\ 

\subsection{Training and Post-Processing Results}

 We attach our \href{https://github.com/chesskarthik01/DG-PPU}{Github Repository} that provides both the training and post-processing algorithm of DG-PPU.  Our post-processing algorithm, when applied to point clouds of partial scans, was successful in getting rid of the false positives and noise of our original 3D point cloud data. We can see this in Fig.2 where we demonstrate side by side the input data (left) and DG-PPU (right) outcome for positions: P1, P2, P3. We’ve circled in red the key areas where DG-PPU outperforms the manual cleaning/de-noising done by our lab technician, Caterina. At P1, we can see that DG-PPU effectively gets rid of points floating between the cluster of patellar and femoral points. Across all positions, we can also see that DG-PPU is effective in making the point clouds smoother. We would like to note that point clouds of partial scans have not been previously seen by our model. This demonstrates that DG-PPU works for non-complete point clouds and has potential use in a clinical setting where thorough scans are not available.\\

We would like to remark that DG-PPU provides an automated post-processing procedure that trains on 3D point clouds extracted from ultrasound, and removes false positives in real-time; this has not been done before. DG-PPU has also got rid of false positives for data sets that it has not trained on. Yet, due to the limited amount of data available, we are uncertain that this will translate across other knee ultrasound images. Further testing with more scans is required to determine whether the post-processing algorithm, without additional training, can be used in clinical settings.

\begin{figure*}
    \centering
   \includegraphics[width=0.7\textwidth ]{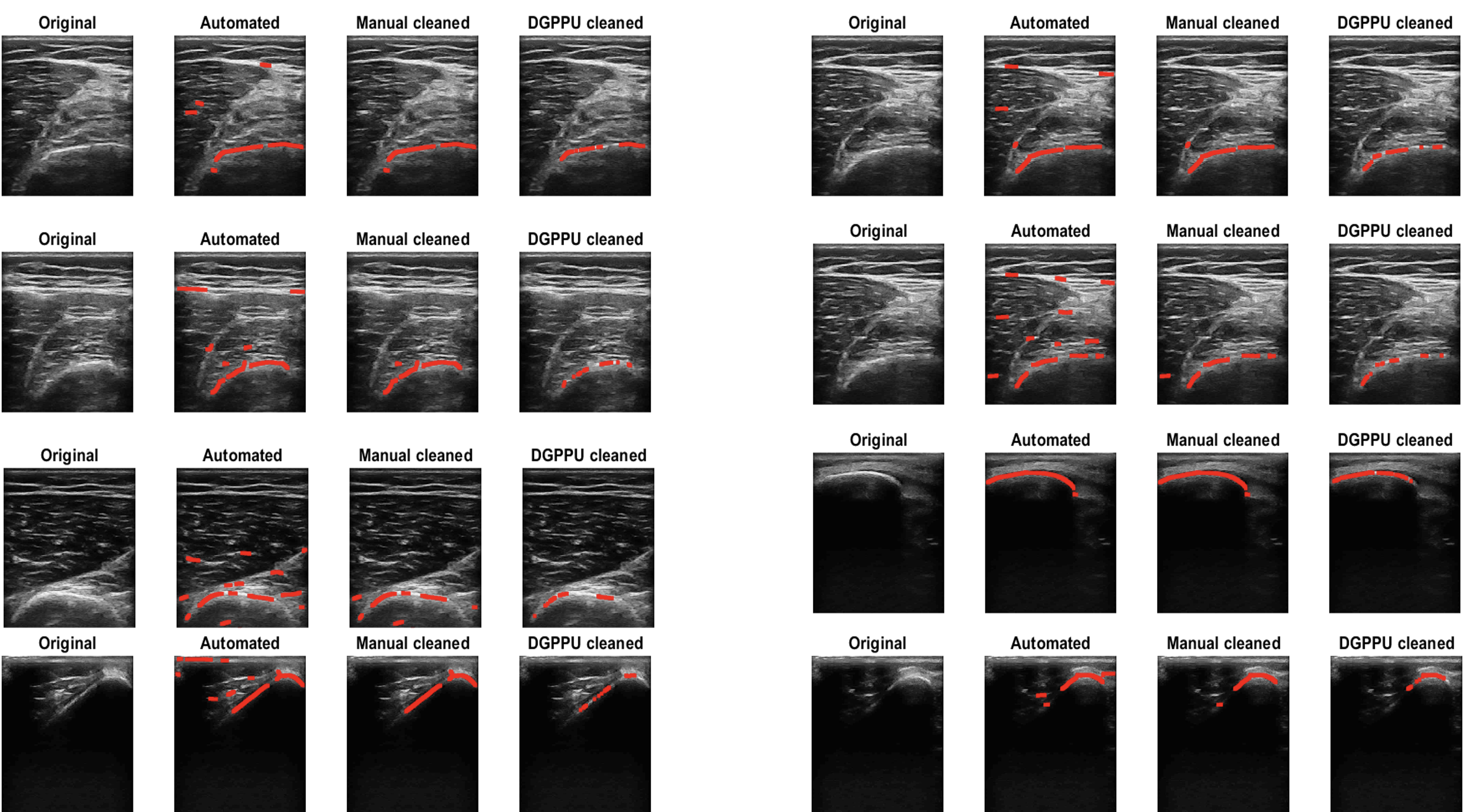}
\caption{DG-PPU compared to medical specialist (manual cleaned) when inverted back to ultrasound scans}
\label{Diagram 4}
\end{figure*}

\section{Evaluation}

 DG-PPU produced smoother point clouds that align more with the 3D anatomy of the knee joint, making them more ideal for 3D position registration. Yet, the following question remains: is the algorithm actually deleting false positives?

As we do not have a ground truth for 3D points, it is challenging to determine what the points we've delete using DG-PPU represent. Medical specialists, however, can look at the original 2D Ultrasound scans and assess whether the part of an image that DG-PPU has deleted contradicts our knowledge of bone anatomy. Hence, we did the inverse operation of CATMAUS to convert the 3D points back to their corresponding frames of ultrasound scans to determine what each point represents. Each coordinate in the point cloud was drawn following a combination of scaling and transformation matrices. We inverted these matrices and their orders to track back to the frame of ultrasound scan that each point represented.\\ 

After comparing the frames containing points we've deleted to the work done by our lab technician Caterina, we demonstrated that our algorithm outperforms manual cleaning.  We refer the reader to Fig.3 for examples of these ultrasound image frames. Lines dotted in red are the parts of the ultrasound scan included in each 3D point cloud. The images labelled \emph{Automated} are the input data and the ones labelled as \emph{Manual Cleaned} are done by Caterina. These images were taken across P1, P2, and P3, representing femoral, patellar, and tibial scans. 

From the inverted ultrasound scans, we also calculated DG-PPU's precision. Precision was measured by excluding frames where DG-PPU either deleted a whole line labelled as bone or left only a single point representing the bone, making it impossible to reconstruct from the output of that frame. Across P1, P2, P3, DG-PPU had precision of \(98.4\%\) (183 out of 186), \(94\%\) (194 out of 200) and \(99.2\%\) (247 out of 249) respectively. Mean precision was \(98.2\%\) overall.\\

It is worthy to highlight that although evaluation was done on 2D ultrasound scans, DG-PPU's primary goal is to post-process the 3D point clouds we extract from the ultrasound scans and aid the CATMAUS point cloud registration pipeline. Evaluating 2D ultrasound scans allow us to point out some mistakes an algorithm can make, which justifies our definition for precision. Segmentation per pixel of 2D ultrasound scans has been studied before using classical segmentation ideas like U-net \cite{U-net}.  We would like to emphasise that such methods are designed for segmentation per pixel in 2D ultrasound scans not 3D point cloud registration in our CATMAUS pipeline.   

\section{Conclusion and further discussion}
 We've established that the dynamical graphs approach allows us to create a novel post-processing algorithm that eliminates false positives and noise of 3D point clouds generated from ultrasounds. DG-PPU produced smooth point clouds that would be more suitable inputs for point cloud registration algorithms and by evaluating the 2D ultrasound scans, we've established that the points the algorithm deleted did not contradict our knowledge of knee bone anatomy. DG-PPU also runs in real-time including both training and post-processing and is robust across different angles of knee joint flexion. In addition, DG-PPU successfully filters partial scans that the algorithm has not trained on, which justifies that DG-PPU has potential clinical usage. However, the limited availability of 3D point cloud data from frame-by-frame ultrasound scans restricts our ability to evaluate DG-PPU. This lack of data makes it challenging to determine if DG-PPU can consistently produce smooth post-processing results as shown in Fig.2. Testing whether this novel algorithm has immediate clinical usage would be an interesting avenue to investigate in the future.\\

Moving back to the point cloud registration pipeline \cite{CATMAUS}, we would like to investigate whether DG-PPU can fit into the pipeline with minimal adjustment. In theory, it would not be surprising if this is achievable, as dynamical graphs allow us to learn the intrinsic geometry that is invariant to knee joint flexion. We also predict that replacing the ICP included in the CATMAUS pipeline with the Deep Closest Point (DCP) algorithm \cite {DCP} would allow us to achieve point cloud registration. DCP makes use of the DGCNN in its pipeline to determine the rigid transformations from one point cloud to another. It would integrate naturally with DG-PPU. Although DG-PPU has corrected a lot of false positives, it may still contain noise for which DCP is robust to compared to its classical counterpart. This further supports our future trajectory of utilising DCP instead of ICP for the point cloud registration of our DG-PPU outputs. If DG-PPU and DCP can successfully achieve point cloud registration in future work, this would enable the development of a novel  patellar tracking assessment system with ultrasound, which currently does not exist.           
\section{Compliance with Ethical Standards}
 This study was conducted in accordance with ethical guidelines and approval from the relevant Institutional Review Board (IRB). Informed consent was obtained from all participants included in the study.
\section{Acknowledgements}
 No funding was received for conducting this study. The authors have no relevant financial or non-financial interests to disclose.

\bibliographystyle{unsrt}    % Choose your desired bibliography style
\bibliography{references}    % Replace 'references' with the name of your .bib file, without the extension
\end{document}